*Studies in Applied Economics*

# FINANCIAL DEEPENING AND ECONOMIC GROWTH IN SELECT EMERGING MARKETS WITH CURRENCY BOARD SYSTEMS: THEORY AND EVIDENCE

Yujuan Qiu

Johns Hopkins Institute for Applied Economics, Global Health, and Study of Business Enterprise

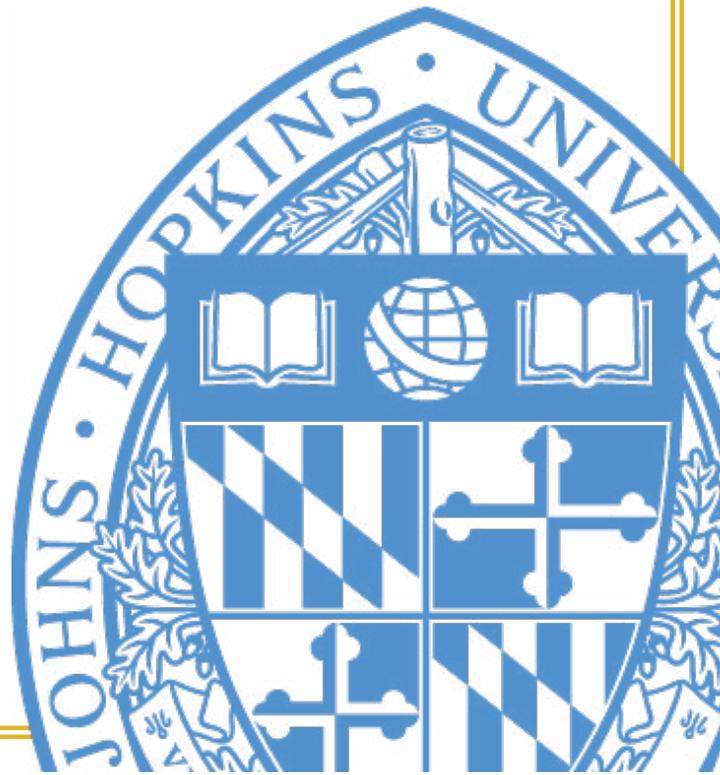

# Financial Deepening and Economic Growth in Select Emerging Markets with Currency Board Systems: Theory and Evidence

By Yujuan Qiu

## About the Series

## About the Author

Yujuan Qiu is a senior at The Johns Hopkins University in Baltimore double majoring in Economics and Applied Math and Statistics. She wrote this paper during her time as an undergraduate researcher for the Institute of Applied Economics, Global Health, and Study of Business Enterprise. She will graduate in May 2018.

## Abstract

This paper investigates some indicators of financial development in select countries with currency board systems and raises some questions about the connection between financial development and growth in currency board systems. Most of those cases are long-past episodes of what we would now call emerging markets. However, the paper also looks at Hong Kong, the currency board system that is one of the world's largest and most advanced financial markets. The global financial crisis of 2008-09 created doubts about the efficiency of financial markets in advanced economies, including in Hong Kong, and unsettled the previous consensus that a large financial sector would be more stable than a smaller one.

## Acknowledgements

I thank Professor Steve H. Hanke and Dr. Kurt Schuler for their guidance and advice.

**Introduction**

Financial development is intertwined with a country's resilience, productivity and growth. The 2008-09 global financial crisis raised doubts about the merits of financial deepening and financial development, given that the crisis originated in advanced economies (AEs), where the financial sector had grown both very large and very complex (Sahay et al. 2015: 5). Debates on the relationship between economic growth and financial development, particularly in relation to emerging markets (EMs), have lasted for years. The role of money and finance in economic growth has been examined and discussed from diverse angles by economists, with little consensus. Joseph A. Schumpeter (1912) and Walter Bagehot (1873) emphasized that the development of the banking system can actively spur innovation and growth by mobilizing savings, managing risks, facilitating transactions, and providing robust funds for productive activities. Raymond Goldsmith (1969) and Ronald McKinnon (1973) also emphasized that an advanced financial structure and a liberalized financial system allow the market to determine its real interest rate, thus exerting a positive effect on the expansion of the economy. On the other hand, Robert Lucas (1988) argued that the role of the financial system is "over-stressed" by economists. Joan Robinson (1952) went further, contending that there is a reverse finance-growth nexus, that is, financial development simply follows economic growth.

Emerging markets are typically characterized by higher financial instability and less developed financial systems than advanced economies. A key element for the stability of an economy is the soundness of its currency. A sound currency is one that is stable, credible, and fully convertible. Stability means that current annual inflation is relatively low, usually in single digits. Credibility means that the issuer creates confidence that it will keep future inflation low. Full convertibility indicates that a currency can buy domestic and foreign goods and services, including buying foreign currencies at market rates without restriction. Unlike advanced economies, emerging markets mostly did not have modern-style central banks before the mid 20$^{th}$ century. Many instead had notes issued by local private commercial banks or by currency boards (Hanke and Schuler 2015: 1). Generally, these systems performed better than the central banks that replaced them. In the worst cases of central bank performance in emerging markets, countries such as Argentina, Bolivia, and Bulgaria have experienced hyperinflations such that their currencies collapsed, leading to sociopolitical upheavals, changes of government and even civil unrest.[1] Historical experience suggests that the establishment of a sound currency in such cases is imperative for promoting durable growth.

Currency boards have generally been successful in providing sound currencies. A currency board is a monetary institution that issues notes and coins (and, in some cases, deposits) fully backed by a foreign "reserve" currency and fully convertible into the reserve currency at a fixed rate and on demand (Hanke and Schuler 2015: 2). The reserve currency is one chosen for its expected relatively good continued performance. Because a currency board has a fixed exchange rate with its reserve currency, an unstable domestic currency is basically replaced by a sound foreign currency and thus stabilizing economy activities and spurring future growth.

---
[1] There have been 57 recorded cases of hyperinflation (Hanke and Bushnell 2016). Almost all have occurred under central banking or direct government issue of currency by the treasury.

Much research has been done to analyze the implementation, feasibility, benefits and improvements of currency board systems in developing economies (for a bibliography, see Gross, Heft, and Rodgers 2012/2013). However, little literature has investigated financial development in currency board systems. As mentioned earlier, money and banks play a crucial role in economic growth and development. Many emerging markets first established and developed advanced financial intermediaries and instruments under currency board systems. However, the finance-growth nexus is still unclear. It is ambiguous whether the establishment of a currency board system encourages financial development, resulting in fast (or at least faster) economic growth.

In this paper, we study whether currency board systems have stimulated higher levels of financial development, and whether currency boards are positively associated with economic growth. We use data from three currency board systems, namely the Hong Kong Exchange Fund (later the Hong Kong Monetary Authority), the Straits Settlements Commissioners of Currency (later the Malayan Currency Board), and the East African Currency Board. Due to limited availability of data, we focus on the six major countries in these currency board systems, namely Hong Kong, Singapore, Malaysia, Uganda, Tanzania, and Kenya. (We omit Brunei in the Malayan Currency Board and a number of countries in the East African Currency Board that joined after the core members.) Because these countries implemented currency board systems at different times and they lasted for different periods, it is difficult to conduct cross-country research by simply comparing financial indicators across common periods. Besides, the availability of data varies, and in most cases, financial data are hard to find. This paper is the first to collect extensive banking data across a number of currency board systems. We start with case studies of selected countries individually, and then discuss general implications for countries that shared currency boards by comparing relatively similar periods and units. Ultimately, we draw some conclusions about trends in different currency board systems.

The main approach is to use ratios of financial assets or liabilities as measures of financial deepening and economic growth. Such measures include the ratio of notes to coins, deposits to currency, and deposits to GDP. Because of differences in the availability of data, the indicators used vary across countries.

For Hong Kong and other cases that have substantial, readily available economic and financial data, we use King and Levine's (1993) methods of measuring economic growth, physical capital accumulation, and economic efficiency improvements. Specifically, we investigate whether higher levels of financial development are spurred by a currency board system, and whether it is significantly and robustly correlated with faster current and future rates of economic growth across countries. If so, it suggests that a currency board not only stabilizes monetary policy but also encourages financial development, and this development has empirical connection with contemporaneous and future long-run economic growth (Levine and Zervos, 1998: 538).

**Methodology**

*Basic Indicators and Ratio Comparison*

One of the most widely used concepts for measuring the development of financial activity is financial deepening. Financial deepening is the increased depth of financial services in an economy, normally measured by the ratio of money supply to GDP or one money supply component to another, such as the ratio of deposits to currency. Specifically, financial deepening indicates an expansion of the financial system, such as more access to financial services, more diverse financial intermediates and instruments, and more availability of risk management for firms and individuals. For currency board systems where data are sparse, we examine certain ratios of more advanced forms of money and credit to more primitive forms. To measure the depth of the financial system and the growth of an economy, we begin our analysis by examining four main financial indicators – monetary base, money supply, deposits, and bank credits and assets – and GDP as a main economic growth indicator. We then compare ratios of those indicators to measure the extent of financial deepening. Table 1 shows details of sub-indicators. However, the analysis of those indicators and ratios of individual countries depends on the availability of data. If data are scarce, only certain aspects of financial deepening can be analyzed.

Individual countries implement currency boards within different historical and economic environments, and they adjust their monetary policies as well as financial regulations accordingly. Simply conducting cross-country research by comparing their indicators during certain periods of time will result in biased and inaccurate estimations that have different scales of time, units of currency, and measurements of GDP.

| DATA SERIES | CALCULATIONS |
|---|---|
| **Monetary Base** | **Money Supply Measures (see Remarks)** |
| Coins in circulation | Monetary base, M0 |
| Notes in circulation | Narrow money, M1 |
| Total currency in circulation | M1 substitute |
| Deposits at the currency board | Broad money, M2 |
| Other or unspecified | M2 substitute |
| Total of all components | Gross domestic product: nominal |
| —Of which: currency held outside of banks | Notes / coins   (code: NCR) |
| Currency held by banks | Deposits (all deposits) / currency   (code: DCC) |
| | M0 / GDP |
| **Deposits** | M1 / GDP   (code: LYY1) |
| Commercial banks: demand deposits | M1 substitute / GDP |
| Commercial banks: time and savings deposits | M2 / GDP   (code: LLY2) |
| Commercial banks: all deposits | Bank credit / GDP |
| Savings banks | Bank assets / GDP |
| Credit cooperatives, etc. | |
| | **Ratio Calculations** |
| **Banking** | Deposits (all deposits) / currency   (code: DCC) |
| Bank foreign assets | M0 / NGDP |
| Bank credit (domestic credit) | M1 / NGDP |
| —Of which: Credit granted to NFPS* | M1 substitute / NGDP |
| Bank assets (= foreign assets + domestic credit) | M2 / NGDP   (code: LLY) |
| | Bank credit / NGDP |
| **Other indicators** | Bank assets / NGDP |
| Nominal gross domestic product (NGDP) | |
| Real GDP growth rate per capita   (code: GDP) | **Other Calculations** |
| Banking offices | Population per bank office   (code: BPP) |
| Population | Private credit / domestic credit   (code: PRIVATE) |
| | Credit to NFPS* / GDP   (code: PRIVY) |

Remarks: *NFPS = nonfinancial private sector. M0 = notes + coins + deposits at currency board; M1 = currency in circulation - currency held by banks + demand deposits; M1 substitute = currency in circulation + demand deposits, where currency held by banks is unknown; M2 = M1 + time deposits or, if no deposit breakdown is available, all deposits; M2 substitute = M1 substitute + time deposits or, if no deposit breakdown is available, all deposits, where currency held by banks is unknown.

**Table 1. Data and Calculations for Financial Ratios**

*Empirical Evidence*

For Hong Kong, where detailed data are available, we adopt and extend the empirical methods of King and Levine (1993). We measure the strength, availability and domestic asset distribution of financial system by using the ratio of liquid liabilities to GDP, the ratio of the number of bank offices to population, the ratio of credits on the nonfinancial private sector to total domestic credit, and the ratio of claims on the nonfinancial private sector to GDP.

We construct four classes of indicators of the depth of financial development. The pioneering work of McKinnon (1973) and Goldsmith (1969) led economists to use the size of the formal financial intermediary sector relative to economic activity to measure financial sector development, or "financial depth" (King and Levine, 1993: 720). One measure of financial depth is the ratio of liquid liabilities of financial system to GDP. We use the ratio of M2 to GDP as an estimator, which we label LLY2. Liquid liabilities are those that can be converted to cash quickly. They consist of currency that is not held by banks, demand deposits of financial institutions at the monetary authority, saving deposits, and time deposits. Parallel to the ratio of M2 to GDP, two sub-indicators measure the financial liquidity and depth of currency and deposits, respectively. The first is the ratio of notes in circulation to coins in circulation, which we term NCR. The second is the ratio of deposits to currency in the circulation, which we term DCC. These two sub-indicators are used to analyze the effect of financial deepening in circulation and deposit sectors.

The second class of indicators of financial development measures the availability and access of financial services for individuals. In our data set for selected countries, the most plausible indicator to measure access to financial services is the number of licensed banks, branches, and offices. Due to the different data availability across countries, we use the number of bank offices, branches, and licensed banks in descending order of priority. The ratio of population to the number of bank offices estimates how many individuals each office is serving on average, which we label BPP. If the data of the number of bank offices is not available, we use the number of branches as a substitute. There are problems with this measure of financial availability: (a) banks are not the only financial intermediaries that provide financial services such as risk management, information acquisition, and monitoring services (King and Levine, 1993: 721); (b) the quantity of banking services alone ignores other key factors such as size and quality; and (c) not all household and individuals are involved in or need financial management. Thus, these measures overestimate the demand for financial services. Nevertheless, by measuring the average volume rate per bank offices, BPP still will provide and complement the analysis of financial availability and depth that can be drawn from LLY.

The third and fourth classes of financial deepening indicators measure domestic credit and assets distribution. One of the most important financial services is to make loans and advances for firms, households, individuals and government. However, a financial system that simply funnels credit to the government or to state-owned enterprises may not be evaluating managers, selecting investment projects, pooling risk, and providing financial services to the same degree as financial systems that allocate credit to the private sector (King and Levine, 1993: 721). Following King and Levine (1993), we construct two indicators to measure credit development. The ratio of credit to the nonfinancial private sector by domestic banks to total domestic credit measures the proportion of credit allocated to private enterprises by the financial system, which we term PRIVATE. The ratio of credit to the nonfinancial private sector to GDP shows how large private credit looms in the economy, and we term this ratio PRIVY. Although the PRIVATE and PRIVY indicators neglect the public sector, and therefore are not complete measurements of financial deepening, they still can illustrate some aspect of financial development and complement the first two indicators in this research.

Corresponding to these four classes of financial indicators, we use the growth rate of real gross domestic product per capita (real GDP per capita) as the measurement of economic growth. In the following sections, we start analyzing the trend of financial indicators in selected countries before and after the installation of their currency boards, and we run regressions of those financial indicators on the growth rate of GDP per capita to investigate correlations between financial development and economic growth. We then perform a cross-country analysis that estimates the general effect of currency boards on financial development by adjusting individual financial indicators to a comparable index.

**Data**

We undertook substantial data collection and digitization from diverse sources. However, in this paper, we only concentrate on select countries that have fuller data. Major sources for the data include B. R. Mitchell's *International Historical Statistics,* recently available in database form; Global Financial Data; World Bank datasets; the International Monetary Fund's International Financial Statistics database; British statistical abstracts of the colonies; local statistical abstracts; Krus and Schuler (2014), incorporated in the Historical Financial Statistics data set; and reports and websites of monetary authorities.

It is difficult to assemble complete datasets for many countries. Banking data, particularly, are often incomplete because many British colonies with currency boards did not require banks to publish statements of local assets and liabilities. The largest banks in most British colonies were London-based organizations with operations in multiple countries, such as today's Standard Chartered Bank. The banks published balance sheets for their global operations, but determining country-level assets and liabilities would involve time-consuming work in bank archives, mainly in Britain, that was not possible within the time and budget for this paper. The underlying data used in this study, and a large amount of other data not used here, are available in an accompanying spreadsheet workbook. Data whose public reproduction is prohibited by their source are available from me for researchers who wish to have the data for personal use in replication of the results here.

**Case Study of Hong Kong**

In the history of Hong Kong, a fixed exchange rate system has been the norm rather than the exception (Chiu n.d: 2). Hong Kong has experienced three currency board periods: from December 1935 to December 1941 (ended by Japanese occupation during World War II), September 1945 to July 1972 (ended by a decision to float the currency to appreciate it during a period of weakness in the pound sterling, then the anchor currency), and from October 1983 to present. Longtime use of a rigid exchange rate anchor has helped make Hong Kong a highly externally oriented economy. For example, visible and invisible total trade accounted for about 300 percent of GDP in 2000. Unlike other countries that adopted currency boards in recent decades as a way to stop hyperinflation, Hong Kong has long historical experience with the system, under which it saw improvement in its social economy and refinement in its financial system. We will focus on Hong Kong's latest currency board period from 1983 to 2002. We stop at 2002 mainly because financial indicators data like coins and notes in circulation only update to 2002 and this period has the best data availability.

Hong Kong, one of the most significant financial centers in the world, has had a successful monetary and financial system compared to other emerging markets. The system developed much differently from the monetary systems of other leading financial centers: Hong Kong has never had a central bank. Before Hong Kong was tied to the U.S. dollar, it adopted a classical British colonial currency board, which itself replaced free banking (fully competitive note issue by banks). The establishment of the Exchange Fund in 1935 represents the beginning of the "sterling exchange era" in Hong Kong, in which Hong Kong and Great Britain shared the same currency. However, the pound sterling was almost continuously weak against the U.S. dollar. The pound was devalued by 30.5 percent on September 19, 1949. On November 18, 1967, the pound was devalued again, by 14.3 percent, which had a critical effect on Hong Kong's economy. Because the value of currencies of other major trade partners such as China and Japan remained stable, Hong Kong's cost of imports rose dramatically. Although negotiations regarding the "Hong Kong Dollar Bond Rate" and the Basel Agreement offered some protection,[2] British monetary instability still had a harmful influence on Hong Kong's economy. When the British government untied the pound from its gold parity and let it float on June 23, 1972, it ended the sterling area — a set of arrangements under which countries that used sterling as their anchor currency allowed capital movements freely among themselves, though not with other countries (Hanke and Culp 2013: 22).

---

[2] As an incentive for Hong Kong to continue to hold sterling reserves, the British government promised to protect the reserves against a further devaluation of the pound sterling with respect to the U.S. dollar.

Britain's floating of the sterling was part of the collapse of the international system of pegged exchange rates that had been agreed at the 1944 Bretton Woods financial conference. Many countries responded with novel policies. Hong Kong at first switched to the U.S. dollar as its anchor currency, then in 1974, when the dollar was suffering from weakness reminiscent of sterling a few years earlier, Hong Kong floated.

Hong Kong at the time had an unusual monetary system that lacked an effective monetary anchor. There was no central monetary authority, so the government did not have control over the supply of interbank liquidity and interest rates to regulate the economy as a whole. Nor did Hong Kong have a currency board system, because no currency or commodity was being used as an exchange rate anchor. The floating exchange rate period was characterized by large swings in the money supply, but compared to many other monetary systems of the era it appeared to work reasonably well until 1983. The United Kingdom and China were negotiating the status of Hong Kong after 1997, when the 99-year British lease on most of the territory of Hong Kong would expire. A senior Chinese official started a panic in Hong Kong when he belligerently stated that China might take Hong Kong back before 1997. The Hong Kong to U.S. dollar exchange rate fell from 7.31 in September to 9.55 in October. To end the panic, the Hong Kong dollar stabilized the exchange rate at 7.8 Hong Kong dollars per U.S. dollar, and returned to a currency board system (Hanke and Culp 2013: 23-25).

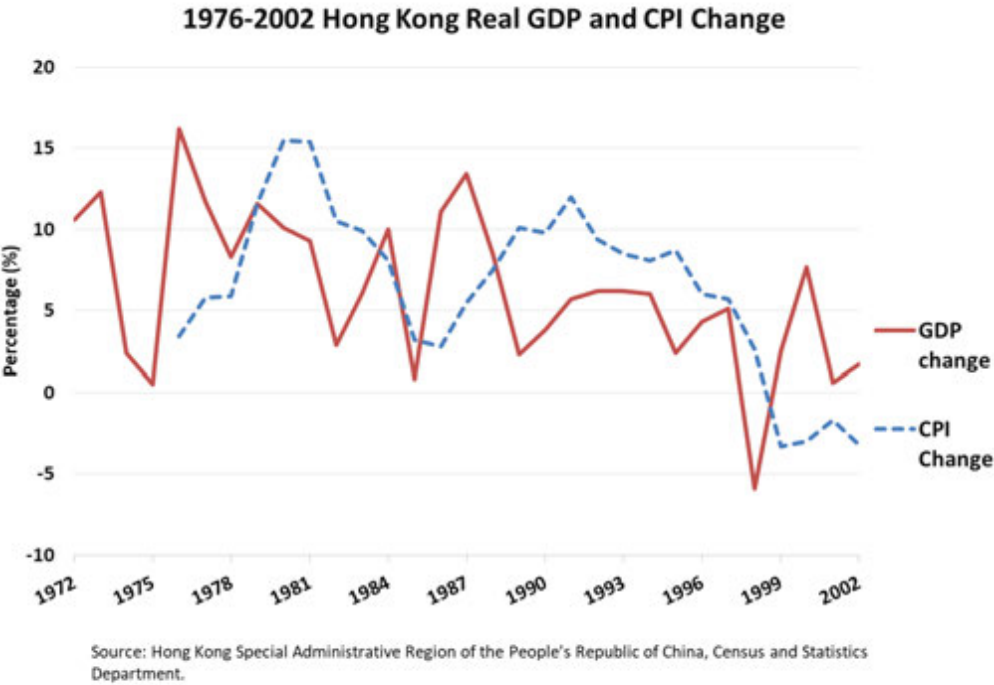

**Chart 1**

As shown in Chart 1, though Hong Kong seemed to recover quickly from the 1973 global oil crisis, achieving real GDP growth of 8-20 percent with around 5 percent inflation from 1976 to 1978, the growth rate kept decreasing. Starting in 1978, real GDP growth dropped from 11.6 percent to 2.9 percent and inflation surged to over 15 percent until 1983. The floating exchange rate regime, public construction projects, and a booming property market created an economic bubble, which burst as the world economy slowed under the influence of recessions in the United States and some other countries.

After the implementation of the currency board system in 1983, there was an immediate rise in real GDP growth, and inflation fell from 15 percent to 4-6 percent within the next five years. Nevertheless, some peculiarities of Hong Kong's currency board system generated complications until reformed in 1988.

Until 1988, the Hong Kong banking system was completely separate from the Exchange Fund, the monetary authority. All payments in the banking system were settled through the Hong Kong Association of Banks (HKAB) clearinghouse rather than through the currency board. The clearinghouse was managed by the Hong Kong and Shanghai Banking Corporation (HSBC) (Hanke and Culp 2013: 25). In other words, the clearing balances and financial transactions of banking system were not shown in the currency board's balance sheet. Furthermore, banks at that time could not directly convert their deposits into banknotes, or vice versa; conversion was indirect and usually worked adequately but was subject to problems under stress (Chiu n.d: 10). Under the interbank clearing and settlement arrangements of the time, there was a lack of rules regarding the provision of liquidity to the banking system and its adjustments to the currency board system. As shown in Chart 1, GDP growth was still volatile. After increasing from 1983 to 1984, it dropped to near zero from 1984 to 1985, then increased sharply from 1985 to 1986 and plummeted again between 1986 and 1988.

The ratio of private credit to total deposits, shown in Chart 2 on the following page, decreased for several years after the adoption of the currency board. An inadequate banking system that restricted financial activity and the defects mentioned in the previous paragraph may explain the decline of bank credits to total deposits, which indicated the need for policy improvements to build confidence in the economy.

Under the interbank clearing and settlement arrangements of the time, there was a lack of rules regarding the provision of liquidity to the banking system and its adjustments to the currency board system. As shown in Chart 1, GDP growth was still volatile. After increasing from 1983 to 1984, it dropped to near zero from 1984 to 1985, then increased sharply from 1985 to 1986 and plummeted again between 1986 and 1988.

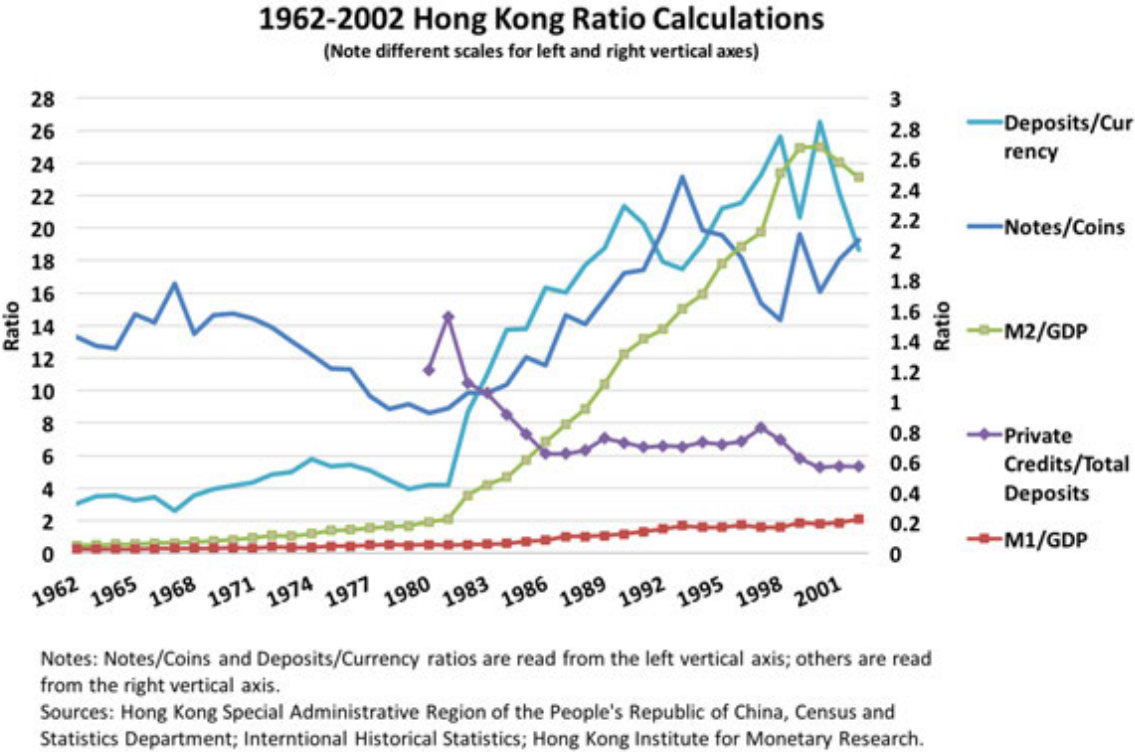

**Chart 2**

The ratio of private credit to total deposits, shown in Chart 2, decreased for several years after the adoption of the currency board. An inadequate banking system that restricted financial activity and the defects mentioned in the previous paragraph may explain the decline of bank credits to total deposits, which indicated the need for policy improvements to build confidence in the economy.

Starting in 1988, the government of Hong Kong launched a series of reforms targeting both the currency board and financial markets. The reforms had three main functions: (i) tightening discipline in the management of interbank liquidity; (ii) setting up a mechanism for the provision of short-term liquidity assistance; and (iii) strengthening the institutional framework for monetary management (Chiu n.d.: 7). The reforms stabilized the economy and improved financial development to some extent by increasing the transparency of financial activities and providing liquidity assistance. From 1988 to 1997, the ratio of private credit to total deposits stayed relatively stable at around 0.7, and other ratio indicators showed steadily increasing trends. Specifically, M2/GDP ratio increased from 0.95 to 2.11 in these nine years. As M2/GDP represents the ratio of the liquid liabilities of the financial system to GDP, its consistent growth implies financial deepening with the implementation of the currency board. Similar trends can be found in the ratio of notes to coins in circulation and the ratio of deposits to currency. Despite a small decrease from 1990 to 1993, the ratio of deposits to currency consistently increased over years. In fact, right after the adoption of the currency board, this ratio grew from around 4 to 10 between 1982 and 1983, and continued rising to 20. The ratio of notes to coins also jumped from 10 to 23 from 1983 to 1993. In 1993 the Exchange Fund and some other government financial functions were merged into the Hong Kong Monetary Authority (HKMA).

Hong Kong's currency board experienced a severe test during the East Asian financial crisis of 1997. As shown in Chart 1, both real GDP growth and inflation turned negative from 1997 to 1999, the first negative growth since 1962. The crisis weakened the Hong Kong dollar dramatically. Collectively selling to the HKMA more Hong Kong dollars than their balances in their clearing accounts, banks faced a liquidity shortage when the foreign exchange transactions had to be settled (Chiu n.d.: 9). Each ratio, though showing the decline from this financial crisis, responded to a different extent, and the trends started at different times. The ratio of notes to coins experienced a major decline three years before the crisis and recovered in 1999 after some amendments to monetary policies were launched. The ratio of deposits to currency, however, was synchronized with financial fluctuations, dropping 5 percentage points between 1997 and 1998. The ratio of private credit to total deposits also decreased around 50 percent during the crisis. Interestingly, the M2/GDP ratio continued rising, jumping 24 percentage points between 1996 to 1998. This continuous increase may indicate that financial activity as a whole played an important role in Hong Kong's economy and steadily boosted growth despite existing financial instabilities.

Besides analyzing the economic activity and financial deepening from ratio indicators, we can also look at a more direct measurement for financial access: the availability of financial services calculated by the ratio of Hong Kong population to total bank offices. As shown in Chart 3 on the following page, after the implementation of the currency board system, financial services developed significantly and stayed at a stable rate in terms of offering services to people per office. From 1962 to 1983, the ratio of population to bank offices plummeted from 16,000 people per office to 3,600 people per office, showing a 4.5 times increase in the availability of financial services. By serving fewer people per office, financial institutions could provide a more systematic, refined and personal services to deepen financial development in Hong Kong. Note that people per banking office was falling before 1983 and then immediately increased from 3,400 to 4,000 people per office at 1983 and continued rising, albeit slowly. This indicates something of a shake-out after the re-introduction of a currency board, perhaps because the fixed exchange rate of a currency board fostered more externally oriented trade and business that required financial services such as transaction, risk management and capital accumulation, or perhaps because of technological changes in the banking industry such as the spread of automatic tellers.

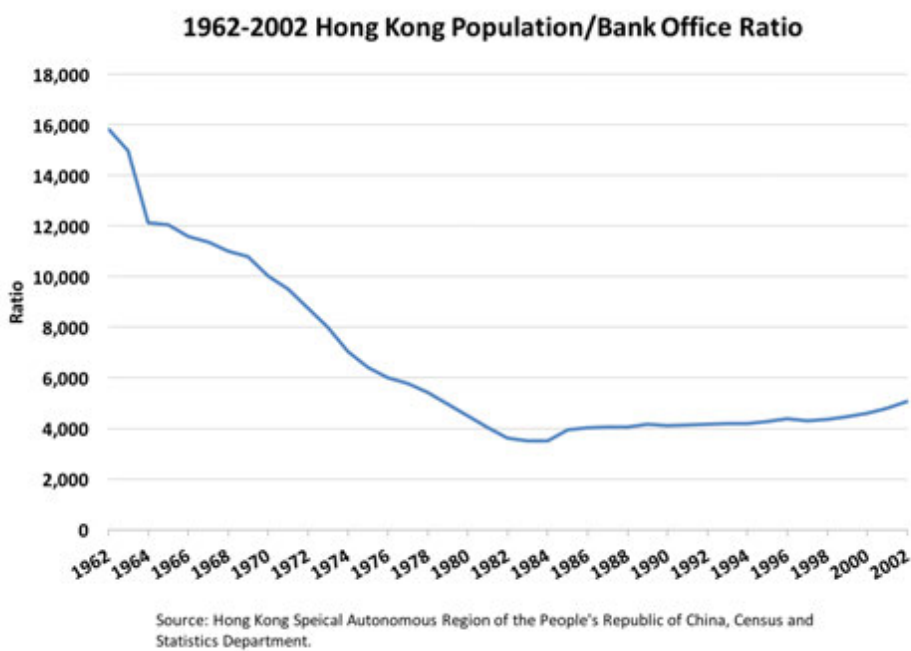

**Chart 3**

According to the analysis of financial ratio calculations and economic development indicators above, financial development was improved by the readoption of a currency board system. However, each financial ratio shows a different degree of effect on economic activities. As the link between financial deepening and economic development becomes clear, it would be helpful if we use financial ratios to analyze and forecast economic activities. Empirical evidence for financial deepening and economic growth are provided as follows. The regression model is:

$$\text{GDP} = \beta_0 + \beta_1 * LYY1 + \beta_2 * LYY2 + \beta_3 * BPP + \beta_4 * PRIVATE + \beta_5 * PRIVY + \beta_6 * NCR + \beta_7 * DCR + U^*$$

Where
GDP = growth rate of real GDP
LYY1 = the ratio of M1 to nominal GDP
LYY2= the ratio of M2 to nominal GDP
BPP = the ratio of population to the number of bank offices
PRIVATE = the proportion of credit allocated to private enterprises by the financial system

PRIVY = the ratio of credits to nonfinancial private sector to GDP
NCR = the ratio of notes in circulation to coins in circulation
DCR = the ratio of deposits to currency in the circulation.
U* = error term

As PRIVATE and PRIVY data are only available from 1979 to 2002 for all variables, we perform regression tests for two periods. (See the accompanying Excel workbook for details.) Regression A includes all available indicators with 23 observations from 1979 to 2002, and Regression C excludes PRIVATE and PRIVY but has 41 observations. As shown in Table 2, Regression A shows that all indicators except PRIVY are statistically significant at the 70 percent level, as their p values are all smaller than 0.3. LYY1, LYY2 and PRIVATE have lower p-values such that we can reject the null hypothesis at the 85 percent significance level. We dropped the statistically insignificant variable, PRIVY, and performed regression B to provide more accurate estimates. The result suggests a linear relationship between financial indicators and GDP growth rate:

$$\text{GDP} = 56.23819 + 219.8881 * LYY1 - 9.829089 * LYY2 - 0.0066327 * BPP - 0.2068211 * PRIVATE - 0.9642619 * NCR + 0.397473 * DCR + U^*$$

Note that four out of six financial indicators in fact have negative coefficients with the GDP growth rate. Only LYY1 and DCR have positive coefficients that are consistent with the finance-growth nexus theory. However, there are two main concerns with the regression models. First, the lack of observations could cause inaccurate estimates. The 23 observations in regression B from 1979 to 2002 years can only provide partial information in that particular period. As mentioned above, the establishment of the currency board in Hong Kong also experienced many complications and economic shocks, and these fluctuations are likely to cause biases. Second, financial indicators are correlated and their correlation can result in multicollinearity in the regression model. Specifically, multicollinearity exists when two or more of the predictors in a regression model are moderately or highly correlated. When it exists, unfortunately, it can wreak havoc on our analysis and weakens the estimated regression coefficient of any one variable. Therefore, regression models that we are using here may involve some bias due to data availability and correlation between financial indicators.

Interestingly, after dropping PRIVATE and adding more observations, the coefficients of some financial indicators show significant change. As shown in Table 2, the new regression, C, shows a linear relationship between GDP and financial indicators:

$$\text{GDP} = 2.301505 + 213.0505 * LYY1 - 15.34502 * LYY2 + 0.001099 * BPP - 0.8939523 * NCR + 0.3112364 * DCR + U^*$$

The coefficient of BPP changes from $-0.0066327$ to $0.001099$, indicating a switch from a negative to a positive correlation with GDP growth rate. While the signs of coefficients remain unchanged, the magnitude of their effects is different between regression B and C. Unlike the implication of finance growth nexus and our earlier analysis of ratio calculations, indicators of financial deepening are in fact negative. Regardless of potential bias, we can interpret the empirical evidence. Regression B suggests that a 1 unit increase of LLY1 and DCR will improve the GDP growth rate by 219.8881 and 0.397473 units, respectively. One unit of increase of LYY2, BPP, PRIVATE and NCR decreases economic growth by 9.829089, 0.0066327, 0.2068211 and 0.9642619, respectively. The R-square is 0.3321, meaning that this model illustrates 33.21 percent of the variation of data. Regression C indicates that a 1 unit increase of LLY1 and DCR improves the GDP growth rate by 213.0505 and 0.3112364 units, respectively, while a 1 unit increase of LYY2 and NCR decreases economic growth by -15.34502 and -0.8939523, respectively. Note that the coefficients of LYY1 and LYY2 have different signs. Although LYY1 is a component of LYY2, LYY2 provides more information and includes time deposits, which increase dramatically after the implementation of the currency board.

We see a mixed result from the regression. Contrary to what the finance growth nexus theory suggests, financial indicators do not have strictly positive relationships with the economic growth rate. Considering the possible biases caused by the lack data availability and multicollinearity, we cannot draw a clear empirical conclusion between financial deepening and economic development. However, the empirical analysis above does provide some new angles and a framework for further studies. If other countries' data are available and allows us to perform more regression tests, we can get more insights for underlying explanations of negative coefficients for these financial indicators. Nevertheless, according to our simple ratio calculation analysis, it is evident that general economic development and financial deepening are synchronized and the implementation of the currency board spurred significant financial progress.

| Dependent variable: GDP | | | | |
|---|---|---|---|---|
| Regressor | Results | A | B | C |
| LYY1 | Coefficient | 227.6041 | 219.8881 | 213.0505 |
| | Standard Error | 133.2964 | 127.8168 | 78.7332 |
| | t-statistics | 1.71 | 1.72 | 2.71 |
| | P>\|t\| | 0.108 | 0.105 | 0.01 |
| LYY2 | Coefficient | -10.1459 | -9.829089 | -15.34502 |
| | Standard Error | 6.615454 | 6.371441 | 5.059007 |
| | t-statistics | -1.53 | -1.54 | -3.03 |
| | P>\|t\| | 0.146 | 0.142 | 0.005 |
| BPP | Coefficient | -0.0069069 | -0.0066327 | 0.001099 |
| | Standard Error | 0.0056259 | 0.0054166 | 0.0004372 |
| | t-statistics | -1.23 | -1.22 | 2.51 |
| | P>\|t\| | 0.238 | 0.238 | 0.017 |
| PRIVATE | Coefficient | -0.2277002 | -0.2068211 | |
| | Standard Error | 0.1273563 | 0.1094728 | |
| | t-statistics | -1.79 | -1.89 | |
| | P>\|t\| | 0.094 | 0.077 | |
| PRIVY | Coefficient | 2.672921 | | |
| | Standard Error | 7.617367 | | |
| | t-statistics | 0.35 | | |
| | P>\|t\| | 0.731 | | |
| NCR | Coefficient | -0.958601 | -0.9642619 | -0.8939523 |
| | Standard Error | 0.716562 | 0.6964734 | 0.4055155 |
| | t-statistics | -1.34 | -1.38 | -2.2 |
| | P>\|t\| | 0.201 | 0.185 | 0.034 |
| DCR | Coefficient | 0.5248003 | 0.397473 | 0.3112364 |
| | Standard Error | 0.4774176 | 0.3016357 | 0.2709365 |
| | t-statistics | 1.1 | 1.32 | 1.15 |
| | P>\|t\| | 0.289 | 0.206 | 0.258 |
| _CONs | Coefficient | 55.47733 | 56.23819 | 2.301505 |
| | Standard Error | 32.25466 | 31.28742 | 4.262536 |
| | t-statistics | 1.72 | 1.8 | 0.54 |
| | P>\|t\| | 0.106 | 0.091 | 0.593 |
| Summary | | | | |
| Years | | 1979-2002 | 1979-2002 | 1962-2002 |
| Adjusted R Square | | 0.31 | 0.3321 | 0.31 |
| Number of Observations | | 23 | 23 | 41 |

**Table 2**

**Malaysia and Singapore's Currency Boards**

*The Straits Settlements, Malaya, and Malaysia*

The Strait Settlements were British territories on the Malayan Peninsula consisting of four individual settlements: Malacca, Dinding, Penang, and Singapore. Singapore established a currency board in 1899. Singaporean currency circulated in the nearby British protectorates on the Malayan Peninsula and to some extent also on the island of Borneo. In 1938 those territories joined with Singapore to create the Malayan Currency Board. All were occupied by Japan during World War II. After the war, the British government broke up the Straits Settlements colony in 1946. The Malayan Union and the Federation of Malaya, which replaced the Malayan Union in 1948, gathered all British territories on the Malayan Peninsula under a single government. Within the Federation, the Malayan states were protectorates of the British Crown, until they achieved independence within the Commonwealth of Nations in 1957. In 1959, Singapore was granted full internal self-government. The Central Bank of Malaysia was established, but it did not undertake activist monetary policy until 1967, when Malaysia, Singapore, and Brunei split their formerly unified currency. In 1963, Singapore, the Federation of Malaya, North Borneo, and Sarawak united to form the independent country of Malaysia. Malaysia's previously inconspicuous central bank began to manage monetary policy in more activist fashion, while Singapore and Brunei each established their own currency boards.

Because of limits of data availability, we only focus on 1949-1967 data for Malaysia, 1963-1973 data for Singapore and 1963-1971 data for both combined. First, concerning nominal GDP growth, we can see from Chart 4 that the early years from 1950 to 1955 performance was poor. Nominal GDP dropped from around 8 billion to 5 billion Malayan dollars as Malaysia's main commodity exports, rubber and tin, experienced a slump after the Korean War. Besides, the gap between real and nominal GDP increased during 1952-1955, indicating the large change of inflation in Malaysian. However, GDP started to increase from 1955 onwards. From 1955 to 1967, nominal GDP increased by around 54 percent. At the same time, financial indicators seemed to reflect this economic growth for some years. As shown in Chart 5, the ratios of bank assets to GDP, M1 to GDP, and M2 to GDP all increase significantly from 1953 to 1955. (Recall that these ratios all concern nominal GDP, not real GDP.) Considering the time lag of the financial effect on the whole economy, this increase of financial indicators could be the effect of the rise of nominal GDP from 1955 onwards.

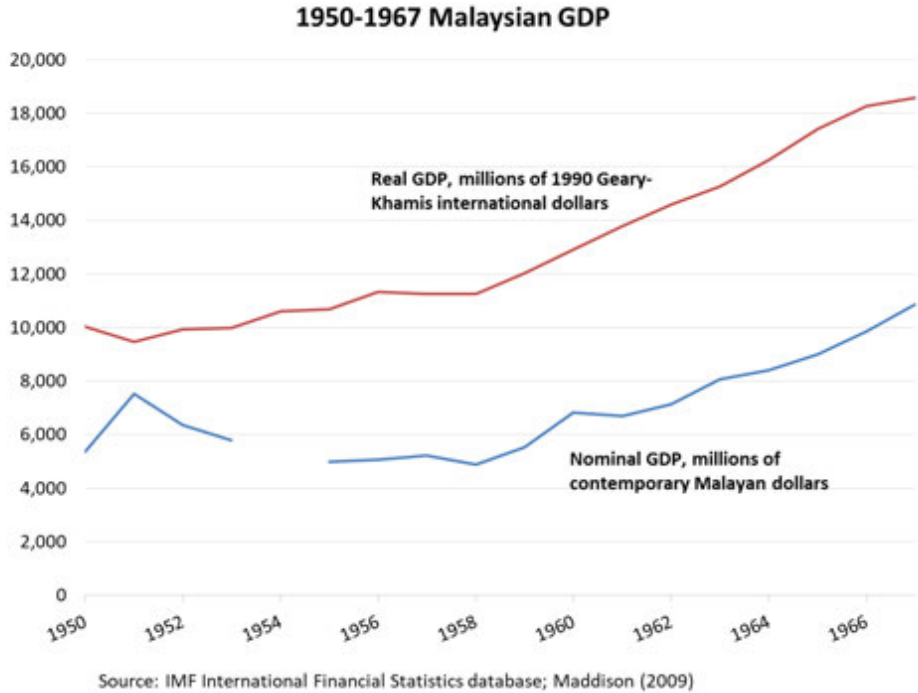

**Chart 4**

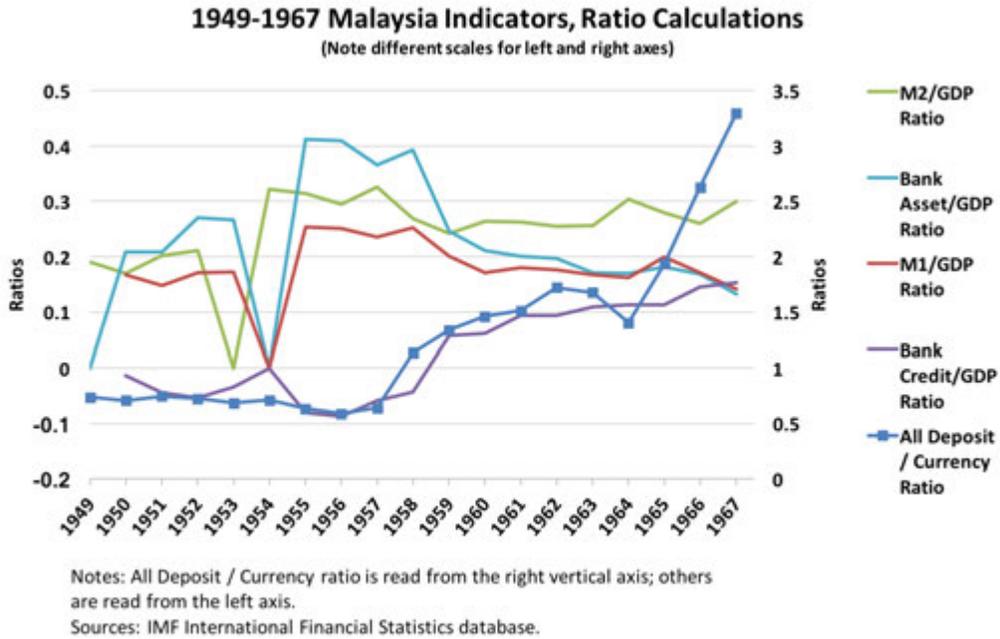

**Chart 5**

Unlike the trends of financial indicators in Hong Kong, Malaysia's financial indicators move together. The ratios of bank credit to GDP and all deposit to currency had similar movements from 1949 to 1964, indicating an internal development of the banking system. The ratios of M2 to GDP, M1 to GDP, and bank assets moved together, reaching a relatively stable growth rate after 1958. These two groups of financial indicators, in fact, could represent two aspects of financial deepening. On the one hand, the ratios of deposit to currency and bank credit to GDP reflect the amount of money that people and investors are willing to put in the bank and the amount of credits that banks managed to lend out, which represents people's confidence in the whole economy. Other ratios, on the other hand, emphasize the liquidity of money. In general, corresponding to the steady growth of nominal GDP, financial indicators also increased from 1959 onwards. Furthermore, a significant increase from 1964 to 1967 for the ratio of deposits to currency may be attributable to the formation of Malaysia in 1963, which by deepening political integration deepened monetary integration.

*Singapore*

Singapore, the most important city in the region, used the Straits dollar between 1845 and 1938. The states of the Malayan peninsula did not issue their own currencies, but rather used Singaporean currency. In 1938 the Malayan states established a joint currency board with Singapore to be able to share in Singapore's profits from currency issue. Singapore joined Malaysia politically in 1963, complementing its existing monetary integration. However, the political union between Malaysia and Singapore broke down soon, and on August 1965, Singapore became independent. In 1967, Singapore and Brunei split from Malaysia in their currencies also, although the Singapore dollar remained equal to the Malaysian ringgit until 1973, when both countries floated their exchange rates. (Brunei kept its currency equal to the Singapore dollar, where it remains today.) Singapore established its own currency board to succeed the Malayan currency board. The currency board system lasted through 1970. At the start of 1971, the Monetary Authority of Singapore, a body with central banking powers, began operations. The Board of Commissioners of Currency Singapore continued to exist as the issuer of notes and coins, and continued to maintain 100% foreign reserve backing for them, but because its liabilities were no longer the sole component of the local monetary base, the Singaporean monetary system as a whole ceased to be a currency board system. We picked years before and after the independence of Singapore, between 1963 and 1973, to analyze the influence of the currency board on Singapore's financial deepening and economic growth.

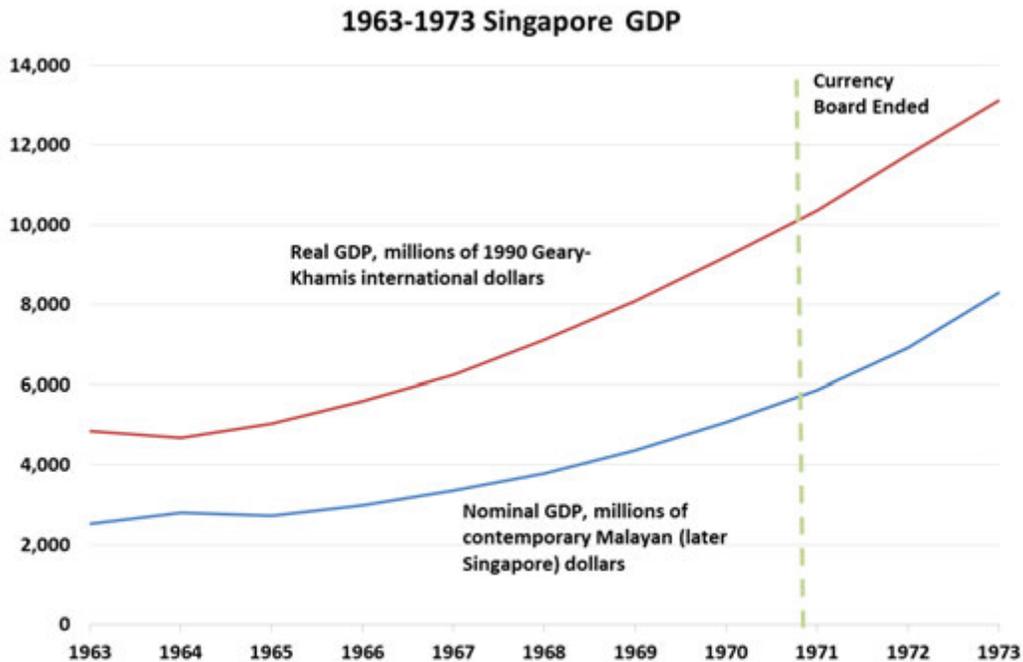

**Chart 6**

As shown in Chart 6, nominal GDP grew steadily but slowly from 1963 to 1966, and it started to increase significantly right after the independence of Singapore. Note that the ratio of all deposits to currency in Chart 7 increased by around 47 percent at the same time that nominal GDP started rising. Besides, the ratios of M2 to GDP and bank assets to GDP experienced a steeper and almost strictly increasing slope from 1966 to 1969. After the announcement of its independence, Singapore, in fact, suffered from high illiteracy rates as well as a high unemployment rate of around 14 percent. Singapore started to expand and stabilize its economy mainly through industrialization. For instance, by the end of 1969, Jurong Industrial Park, one of the earliest industrial areas in Singapore, cost Singapore $45.7 million to set up for 153 fully-functioning factories with a total of 14.78 square kilometers. With effective government policies to tackle economic problems and stimulate financial activities, Singapore's economy continued to grow and attract foreign investments. As shown in Chart 7, the ratio of bank credit to GDP increased by around 30 percent from 1971 to 1973, as well as the ratio of M2 to GDP and M1 to GDP. These continuously increasing trends for financial indicators show that financial development played a crucial role in economic growth.

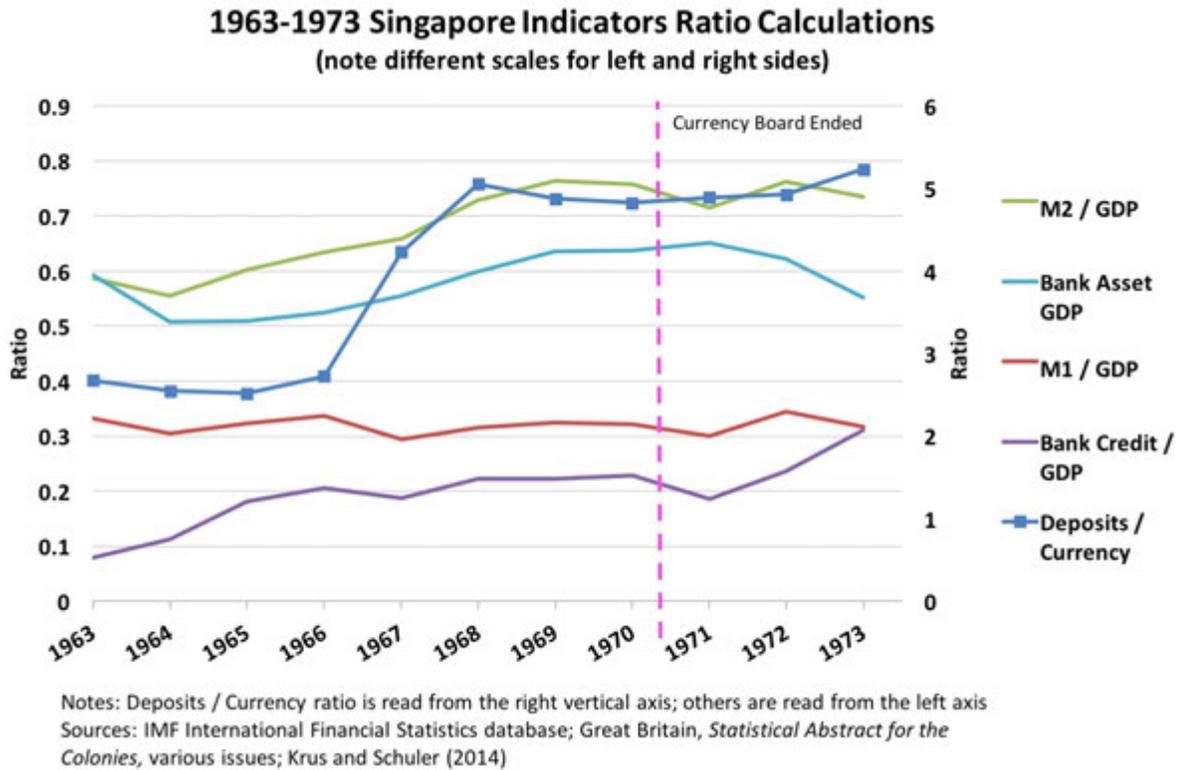

**Chart 7**

*Summary of Ratios under the Currency Board in Singapore and Malaysia*

We have seen that financial indicators reflect economic development and institutional changes in both Singapore and Malaysia during their period of a common currency board. Charts 8 and 9 show the development of various financial ratios.

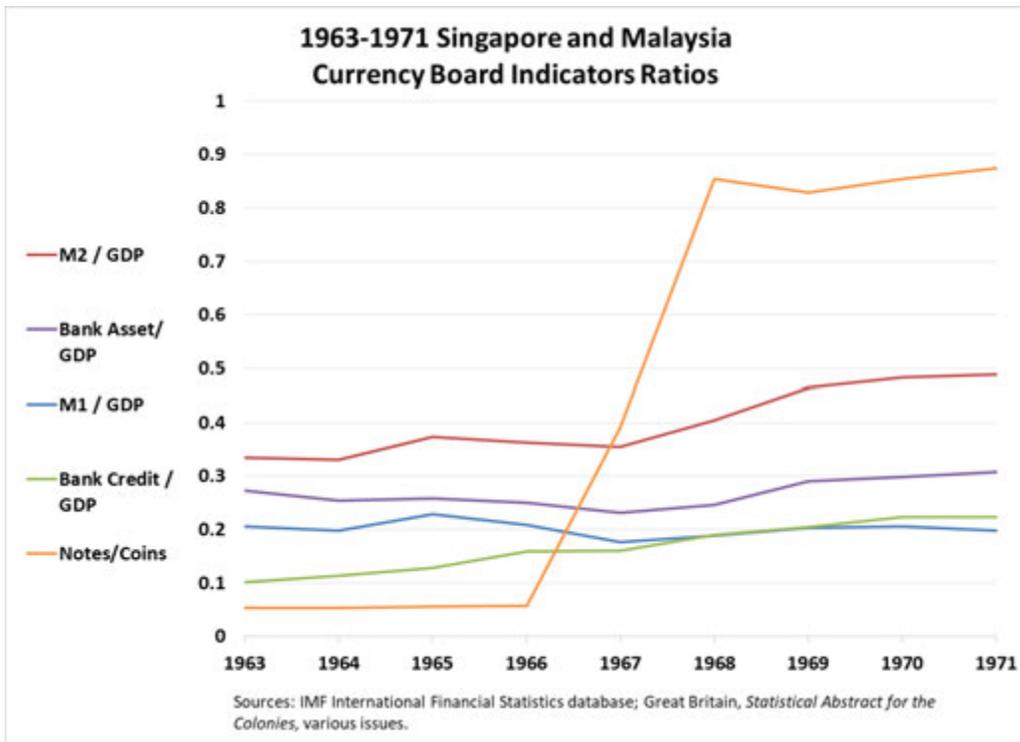

**Chart 8**

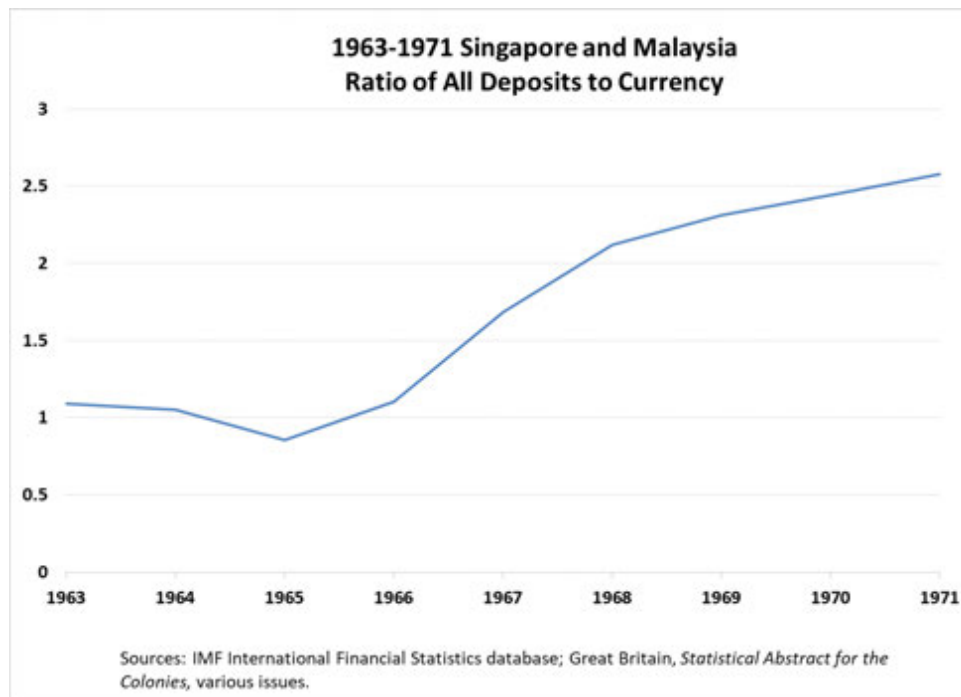

**Chart 9**

Notice that the independence of Singapore in 1966 causes a significant increase of the ratio of notes and coins. As mentioned above, Singapore started to issue its own coins and notes in 1967. In general, all of these financial ratios raised significantly with economic and the growth continued during the implementation of currency board.

**East African Currency Board**

The East African Currency Board (EACB) was established in December 1919 as a successor to the currency board of Kenya. The EACB was a joint currency board for Kenya, Uganda and Tanganyika (the mainland of present-day Tanzania). Tanzania was originally part of German East Africa, but was reallocated to British administration by the League of Nations after the German defeat in World War I. Whereas the profits of the Kenyan currency board belonged to the Kenyan government alone, the EACB split its profits among its members. Zanzibar joined in 1936, relinquishing its own currency board. During World War II the EACB expanded its operations to Somalia, Ethiopia, and Eritrea after British forces defeated Italian forces in 1942 and 1943 (Ethiopia left the EACB and started issuing its own currency in 1945). The EACB even extended across the water to the British colony of Aden (now southern Yemen) for several years. However, Kenya, Tanzania, and Uganda remained the core members of the board.

The EACB, like other British currency boards, operated as a sort of automatic money changer. It issued legal tender locally on demand against payment of sterling in London and redeemed the local currency on demand by paying out sterling (Kratz 1966). It did not control the quantity of currency in circulation through independent monetary policies, but it was responsible for printing notes, minting coins and fixing the denomination of coins and notes. The East African currency unit, the shilling, was equal to the British shilling, which is to say that it was 1/20 of a pound sterling. The East African pound, a widely used informal unit of account, was equal to the pound sterling. The EACB, like many 20$^{th}$ century currency boards in British colonies, was modeled on the West African Currency Board (WACB), which had been established in 1912 by the British. The main purpose of establishing currency boards in colonies was to fix the exchange rate of the colonies to the metropole to eliminate foreign exchange risk in trade, in addition to gaining full control over the monetary policy of the colonies.

Before the implementation of the EACB, the Indian rupee was the main currency or anchor currency of British protectorates and colonies in East Africa. At the outset, the EACB faced many difficulties in its attempt to replace the Indian rupee and, later, coins of Indian standard issued by the German government (Kratz 1966). The former circulated in Kenya and Uganda and the latter in Tanganyika. These coins were replaced because the members of the EACB wanted to capture the profits from issuing coins. But whereas the WACB had been able to repatriate the British silver coins formerly circulating in West Africa at their face value, the EACB was not successful in making similar arrangements with the Indian and German governments. The price of silver was high during the period when the EACB bought old silver coins but then plunged. The EACB had no choice but to sell the retired coins for sterling at their bullion value, which resulted in a loss of more than 1.75 million East African shillings (Kratz 1966).

Despite initial difficulties in its implementation, the available data suggest that the EACB did improve financial development in East African colonies. Significant differences in the availability of data across EACB countries make it difficult to compare financial indicators of the three countries individually, though for some indicators aggregate statistics exist. Therefore, we will analyze EACB financial indicators across different periods of years that have greatest availability.

We gathered data for coins and notes in circulation for the EACB from 1921 to 1971 and calculated the ratio of notes to coins as shown in Chart 10. Although our data continue to 1971, EACB in fact ended in 1966. After the creation of the currency board system, the ratio was around 0.7 and never exceeded 1, indicating that notes were slightly fewer than coins until 1939, when World War II began. Wartime inflation stimulated the demand for notes, thus leading notes to increase to two times more than coins. The amount of notes in circulation rose about five-fold from 1943 to 1947, as Chart 11 shows. Another reason for the dramatic increase of notes was the temporary spread of the EACB's territory to former Italian colonies. After 1947, there was then a postwar pause as Ethiopia, Eritrea, and Italian Somaliland (now southern Somalia) replaced EACB currency with local currency, reducing the quantity of notes by around 18 percent.

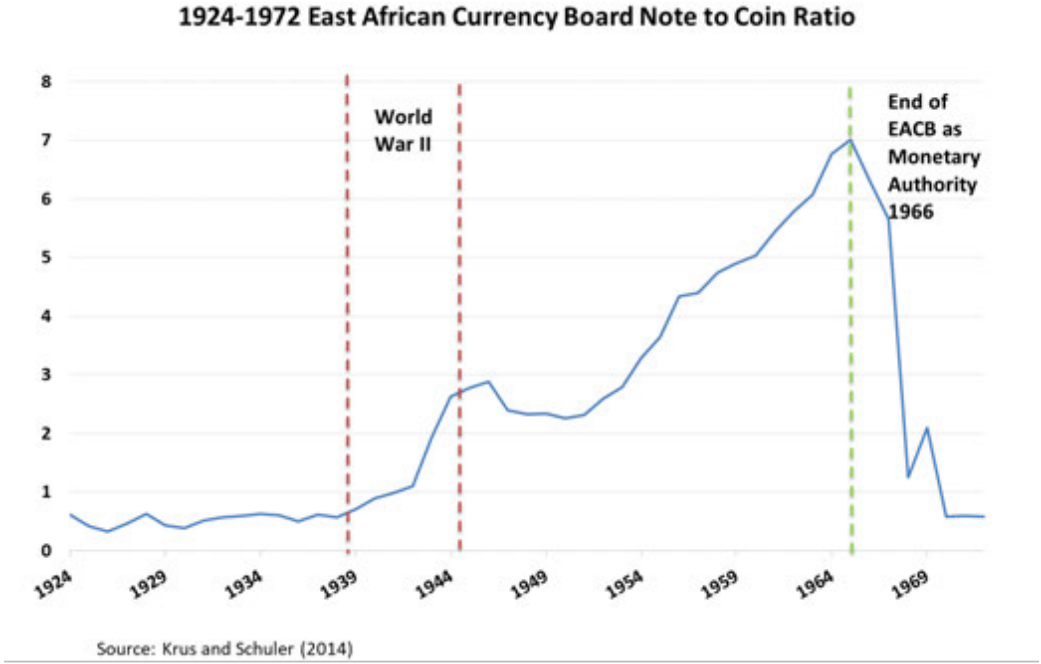

**Chart 10**

However, in the early 1950s note circulation again rose sharply as East Africa grew and financial development increased. Also in this period, the EACB gained some discretionary powers. In December 1954, the British Secretary of State for the Colonies announced that EACB could extend credits to Government of the constituent territories by holding government securities with the limits to £10 million. In 1957, the regulation was amended to increase to limit to £20 million, following with the authorization of acquiring East African Treasury bills in 1959. Note that the circulation dropped dramatically in 1967 when Kenya, Uganda, and Tanzania replaced the EACB with national central banks issuing separate national currencies.

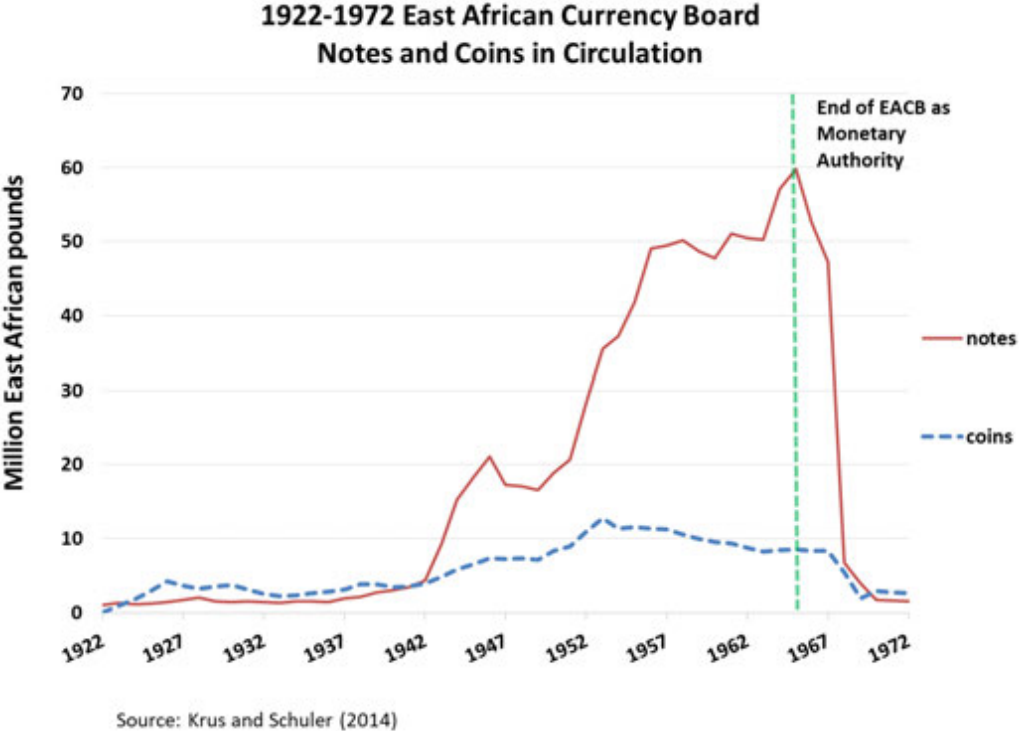

**Chart 11**

Not only did the implementation of the currency board enhance the circulation of coins and notes, but it stimulated or at least allowed the growth of savings deposits. Charts 12 and 13 show that savings bank deposits in all three countries show a similar upward trend during the currency board period. Chart 12 shows that all rose in the late 1920s, slowed or stagnated during the Great Depression, then resumed growth. In Chart 13, savings deposit grew until 1955. This emerging diversity of asset forms may have encouraged investors and savers to put money in other investments with relatively high returns, thus slightly decreasing the volume of saving deposits after 1955.

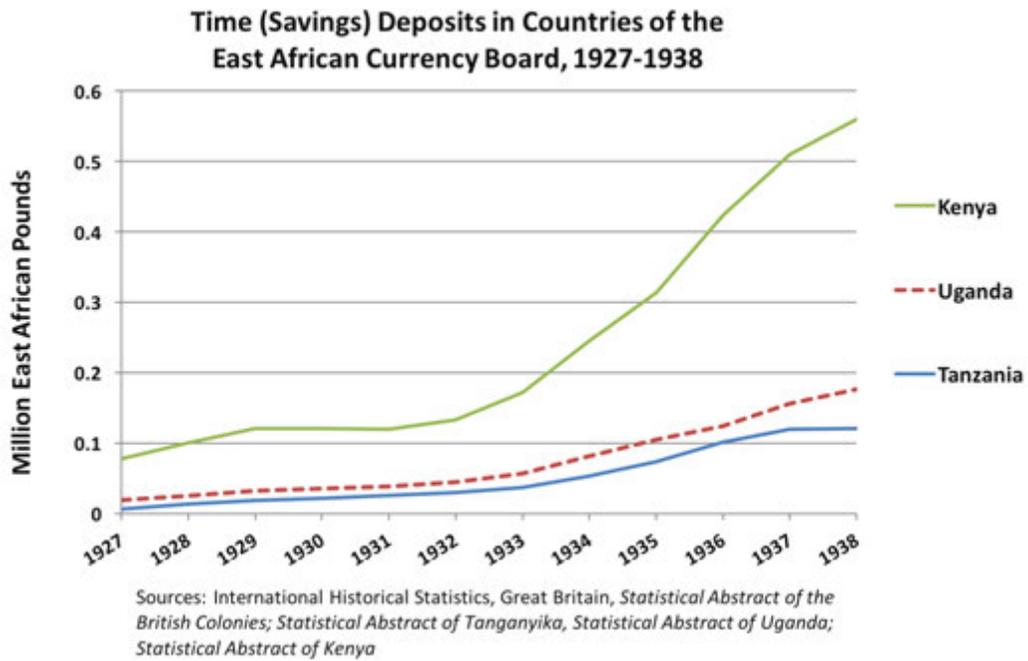

**Chart 12**

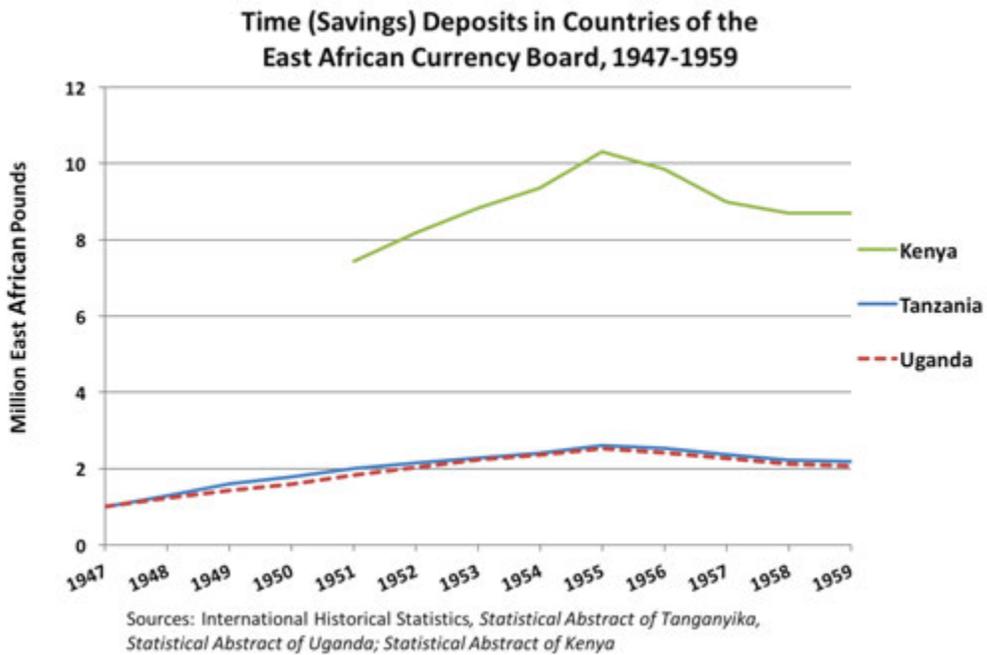

**Chart 13**

**Conclusion**

This paper has investigated financial deepening and economic growth in three currency board systems: Hong Kong, Singapore / Malaysia, and East African. The novelty of this paper is its collection of the currency boards' financial data as well as their corresponding ratio calculations, which were digitized and analyzed systematically for the first time. We conclude that there is an increase in financial deepening along with the economic growth during the years of the adoption of each currency board system analyzed. Financial deepening can bring important benefits to emerging market economies. Currency board systems seem to help stabilize the economy and spur improvement of both monetary and financial systems. At least, up to a point they do: our empirical analysis of Hong Kong, which had by far the most advanced financial system of those studied here, shows an ambiguous causal relationship between financial development and economic growth. Further studies could be carried out by controlling the effects of a currency board to compare their financial growth to countries that did not have currency board systems.